# Separability of pure *n*-qubit states: two characterizations

Philippe Jorrand and Mehdi Mhalla

*Leibniz Laboratory, IMAG Institute*
*Grenoble, France*

Philippe.Jorrand@imag.fr, Mehdi.Mhalla@imag.fr

ABSTRACT. Given a pure state $|\psi_N\rangle \in \mathcal{H}_N$ of a quantum system composed of *n* qubits, where $\mathcal{H}_N$ is the Hilbert space of dimension $N = 2^n$, this paper answers two questions: what conditions should the amplitudes in $|\psi_N\rangle$ satisfy for this state to be separable (i) into a tensor product of *n* qubit states $|\psi_2\rangle_0 \otimes |\psi_2\rangle_1 \otimes ... \otimes |\psi_2\rangle_{n-1}$, and (ii), into a tensor product of 2 subsystems states $|\psi_P\rangle \otimes |\psi_Q\rangle$ with $P = 2^p$ and $Q = 2^q$ such that *p+q=n*? For both questions, necessary and sufficient conditions are proved, thus characterizing at the same time families of separable and entangled states of *n* qubit systems. These conditions bear some relation with entanglement measures, and a number of more refined questions about separability in *n* qubit systems can be studied on the basis of these results.

## 1. INTRODUCTION

Most major results in quantum information processing and communication do rely upon quantum entanglement as their key ingredient: quantum dense coding [1] and teleportation [2], quantum algorithms [3]-[7], quantum cryptography [8], multiparty quantum computation [9][10]. For historical and foundational reasons, which date back to the controversy raised by Einstein, Podolsky and Rosen in 1935 [11], followed in 1964 by Bell's inequalities [12] and their violations in 1982 [13], entanglement has been most extensively studied in the simplest situation where it may occur, namely within quantum systems composed of two two-level subsystems. However, compared with the abundant literature about entanglement in two qubit systems (see for example [14]-[16] and the papers they refer to), there still are rather few in depth and detailed studies of more general situations, where more than two subsystems may be entangled. A physicist's reason is, certainly, the extreme difficulty in preparing and maintaining entangled states of more than two quantum subsystems. Because of this, research in quantum information processing and communication is sometimes considered optimistic.

This paper is on the optimistic side: entanglement is studied within systems composed of *n* qubits, with $n \geq 2$. A few other recent papers, among them [17][18], deal with similar situations. Some of the results in [17] will be commented upon later because they are not unrelated with the results presented in this paper.

Philippe Jorrand and Mehdi Mhalla

DEFINITION 1. *Full separability*. Given a pure state $|\psi_N\rangle \in \mathcal{H}_N$ of a quantum system composed of $n$ qubits, where $\mathcal{H}_N$ is the Hilbert space of dimension $N = 2^n$, $|\psi_N\rangle$ is *fully separable* iff it can be factorized into a tensor product of $n$ qubit states, each of them in $\mathcal{H}_2$:

$$|\psi_N\rangle \text{ is fully separable} \Leftrightarrow \exists |\psi_2\rangle_0, |\psi_2\rangle_1, ..., |\psi_2\rangle_{n-1} \in \mathcal{H}_2 : |\psi_N\rangle = |\psi_2\rangle_0 \otimes |\psi_2\rangle_1 \otimes ... \otimes |\psi_2\rangle_{n-1}$$

DEFINITION 2. *p-q separability*. Given integers $p$ and $q$ such that $p + q = n$, $|\psi_N\rangle$ is *p-q separable* iff it can be factorized into a tensor product of a subsystem state $|\psi_P\rangle \in \mathcal{H}_P$ with a subsystem state $|\psi_Q\rangle \in \mathcal{H}_Q$ where $P = 2^p$ and $Q = 2^q$ (i.e. the two subsystems are composed of $p$ and $q$ qubits respectively):

$$|\psi_N\rangle \text{ is } p\text{-}q \text{ separable} \Leftrightarrow \exists |\psi_P\rangle \in \mathcal{H}_P, |\psi_Q\rangle \in \mathcal{H}_Q : |\psi_N\rangle = |\psi_P\rangle \otimes |\psi_Q\rangle$$

Clearly, full separability implies *p-q* separability, but the converse is not true, since *p-q* separability does not tell anything about the separability properties of the subsystems, i.e. about the presence or absence of entanglement within each of the separated subsystems. It is also useful to notice that *p-q* separability relies on a proper ordering among the qubits: if the *n* qubits are numbered from 0 to $n-1$, $|\psi_P\rangle$ is indeed the state of the subsystem composed of the qubits numbered from 0 to $p-1$, whereas $|\psi_Q\rangle$ is the state of the subsystem composed of qubits $p$ to $n-1$. As a consequence a reordering among the qubits will in general be required for *p-q* separability to appear provided that $|\psi_N\rangle$, as modified after reordering the qubits, is indeed *p-q* separable.

Both full separability and *p-q* separability are invariant under unitary operations local to any of the subsystems represented by their states as written in the tensor product. In [17], entanglement and separability are studied by making explicit the consequences of such invariance properties: this is done first in the case of systems composed of two *N*-dimensional subsystems, then in the case of systems composed of three *N*-dimensional subsystems, after which a generalization to *M N*-dimensional subsystems is briefly presented. The approach taken here is different, although the questions addressed are quite close. Two questions are answered, both about $|\psi_N\rangle$, i.e. starting as initial data with the raw knowledge of the $2^n$ amplitudes of a pure state:

- Is $|\psi_N\rangle$ fully separable?

- Given *p* and *q*, is $|\psi_N\rangle$ *p-q* separable?

The first question is treated in part 2 of the paper and the second question in part 3. In both cases, necessary and sufficient conditions are proved according to which $|\psi_N\rangle$ is indeed (fully, *p-q*) separable. The negations of these conditions thus also characterize corresponding families of entangled states. The last section of the paper draws attention to more refined questions which are worth studying further on the basis of these results about separability and entanglement in *n* qubit systems.

## 2. FULL SEPARABILITY

Let $\alpha_i, 0 \leq i \leq N-1$, $N = 2^n$, be the amplitudes of $|\psi_N\rangle \in \mathcal{H}_N$, with $\sum_{i=0}^{N-1} |\alpha_i|^2 = 1$:

$$|\psi_N\rangle = \sum_{i=0}^{N-1} \alpha_i |i\rangle$$

DEFINITION 3. *Pair product invariance.* Pair product invariance is a property of $|\psi_N\rangle$ defined as follows:

$|\psi_N\rangle$ is pair product invariant $\Leftrightarrow \forall k \in [1, n], \forall i \in [0, K-1]: \alpha_i \alpha_{K-i-1}$ is constant where, for any integer $k$, $K$ denotes $2^k$.

The pair product invariance of $|\psi_8\rangle$, for example, is visualized in figure 1, where the products of pairs of amplitudes linked by arrows with the same color (i.e. with the same *k*) all have the same value if $|\psi_8\rangle$ is indeed pair product invariant:

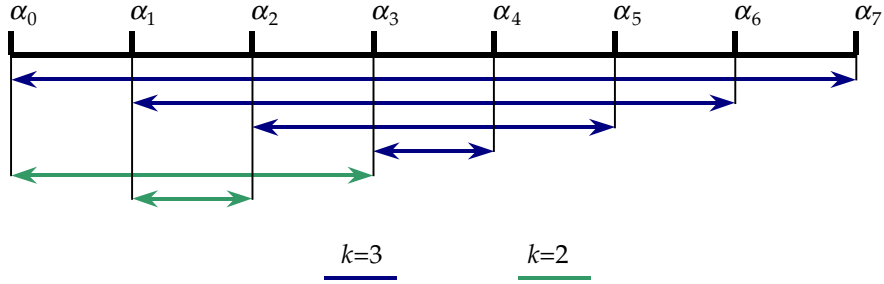

Fig. 1: Pair product invariance of $|\psi_8\rangle$

If $|\psi_N\rangle$ is pair product invariant, many other equalities among products of pairs of amplitudes are satisfied. The definition of pair product invariance provides a minimal set of such equalities from which all others can be obtained.

2.1. SIMPLE CASE: NO AMPLITUDE IS ZERO

Among the consequences of pair product invariance, equalities of the form $\alpha_{2i-1}\alpha_{2i} = \alpha_{2i-2}\alpha_{2i+1}$, with $1 \leq i \leq 2^{n-1} - 1$, are of special interest for proving the conditions of full separability of $|\psi_N\rangle$. The following lemma shows that these equalities indeed hold whenever $|\psi_N\rangle$ is pair product invariant, provided that no amplitude of $|\psi_N\rangle$ is zero:

LEMMA 1. Let $|\psi_N\rangle \in \mathcal{H}_N$ be a state with no zero amplitude. Then:

$|\psi_N\rangle$ is pair product invariant $\Rightarrow \forall i, 1 \leq i \leq 2^{n-1} - 1: \alpha_{2i-1}\alpha_{2i} = \alpha_{2i-2}\alpha_{2i+1}$

Philippe Jorrand and Mehdi Mhalla

PROOF. By induction on index $i$.

Base case: for $i = 1$, $|\psi_N\rangle$ is pair product invariant $\Rightarrow$ $\alpha_1 \alpha_2 = \alpha_0 \alpha_3$

Induction step: assume the property holds for indices 1, 2, …, $i$, with $2(i+1) < N-1$. It is always possible to find an integer $k < n-1$ such that $K < i+1 \le 2K$ (remember: $K = 2^k$). This implies that $2K - (i+1) \le K < i+1$, so that the induction hypothesis holds for index $2K - (i+1)$:

$$\alpha_{2(2K-i)-3}\alpha_{2(2K-i)-2} = \alpha_{2(2K-i)-4}\alpha_{2(2K-i)-1} \tag{1}$$

In addition, noticing that $4K \le N$ because of $k < n-1$, the pair product invariance of $|\psi_N\rangle$, means that:

$$\alpha_{2i+1}\alpha_{4K-1-(2i+1)} = \alpha_{2i}\alpha_{4K-1-2i}$$

and $\quad \alpha_{2i+2}\alpha_{4K-1-(2i+2)} = \alpha_{2i+3}\alpha_{4K-1-(2i+3)}$

Combining these two equalities and reformulating the indices give:

$$\alpha_{2i+1}\alpha_{2i+2}\alpha_{2(2K-i)-3}\alpha_{2(2K-i)-2} = \alpha_{2i}\alpha_{2i+3}\alpha_{2(2K-i)-4}\alpha_{2(2K-i)-1}$$

From (1), and because no amplitude is zero, this can be simplified to:

$$\alpha_{2(i+1)-1}\alpha_{2(i+1)} = \alpha_{2(i+1)-2}\alpha_{2(i+1)+1} \qquad \text{c}$$

For example, in the case of $|\psi_8\rangle$, lemma 1 distinguishes three equalities among those deducible from pair product invariance. They are pictured each with a different color in figure 2:

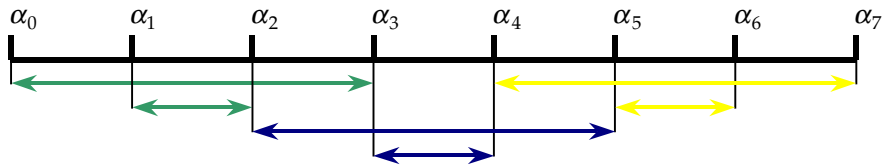

Fig. 2: Lemma 1 equalities from pair product invariance of $|\psi_8\rangle$, assuming no zero amplitude in $|\psi_8\rangle$

Saying that $|\psi_N\rangle$ is fully separable is saying that there exist qubit states $|\psi_2\rangle_0, |\psi_2\rangle_1, \ldots, |\psi_2\rangle_{n-1}$ such that $|\psi_N\rangle = |\psi_2\rangle_0 \otimes |\psi_2\rangle_1 \otimes \ldots \otimes |\psi_2\rangle_{n-1}$. If the amplitudes of each $|\psi_2\rangle_j \in \mathcal{H}_2$, $0 \le j \le n-1$, are:

$$|\psi_2\rangle_j = \begin{pmatrix} \gamma_{0,j} \\ \gamma_{1,j} \end{pmatrix}$$

then:

$$|\psi_N\rangle = \begin{pmatrix} \alpha_0 \\ \alpha_1 \\ \vdots \\ \alpha_{N-1} \end{pmatrix} = \begin{pmatrix} \gamma_{0,0} \\ \gamma_{1,0} \end{pmatrix} \otimes \begin{pmatrix} \gamma_{0,1} \\ \gamma_{1,1} \end{pmatrix} \otimes \ldots \otimes \begin{pmatrix} \gamma_{0,n-1} \\ \gamma_{1,n-1} \end{pmatrix}$$

THEOREM 1. Let $|\psi_N\rangle \in \mathcal{H}_N$ be a state with no zero amplitude. Then:

$$|\psi_N\rangle \text{ is fully separable} \quad \Leftrightarrow \quad |\psi_N\rangle \text{ is pair product invariant}$$

PROOF. Case $\Rightarrow$: by induction on $n$, the number of qubits.

Base case: for $n=2$, if $|\psi_4\rangle$ is separable, then the well known equality $\alpha_1 \alpha_2 = \alpha_0 \alpha_3$ is satisfied, which is identical to pair product invariance of $|\psi_4\rangle$.

Induction step: assume the $\Rightarrow$ property holds for $2, 3, \ldots n$ qubits. Let $|\psi_{2^{n+1}}\rangle = |\psi_{2N}\rangle$ be a state with no zero amplitude and fully separable into a product of its $n+1$ component qubit states:

$$|\psi_{2N}\rangle = \begin{pmatrix} \beta_0 \\ \beta_1 \\ : \\ \beta_{2N-2} \\ \beta_{2N-1} \end{pmatrix} = \begin{pmatrix} \alpha_0 \\ \alpha_1 \\ : \\ \alpha_{N-1} \end{pmatrix} \otimes \begin{pmatrix} \delta_0 \\ \delta_1 \end{pmatrix} = \begin{pmatrix} \alpha_0 \delta_0 \\ \alpha_0 \delta_1 \\ : \\ \alpha_{N-1} \delta_0 \\ \alpha_{N-1} \delta_1 \end{pmatrix} \quad \text{where} \quad \begin{pmatrix} \alpha_0 \\ \alpha_1 \\ : \\ \alpha_{N-1} \end{pmatrix} = \begin{pmatrix} \gamma_{0,0} \\ \gamma_{1,0} \end{pmatrix} \otimes \begin{pmatrix} \gamma_{0,1} \\ \gamma_{1,1} \end{pmatrix} \otimes \ldots \otimes \begin{pmatrix} \gamma_{0,n-1} \\ \gamma_{1,n-1} \end{pmatrix} = |\psi_N\rangle$$

With $k \leq n+1$, if $i$ is even, $K-i-1$ is odd, and vice-versa. Then:

$$\beta_i \beta_{K-i-1} = \delta_0 \delta_1 \alpha_{\lfloor i/2 \rfloor} \alpha_{\lfloor (K-i-1)/2 \rfloor}$$

has the property that $\lfloor i/2 \rfloor + \lfloor (K-i-1)/2 \rfloor = \frac{K}{2} - 1$. Therefore, for a given $k$, by induction hypothesis on $|\psi_N\rangle$, $\alpha_{\lfloor i/2 \rfloor} \alpha_{\lfloor (K-i-1)/2 \rfloor}$ is independent of $i$ and so is also $\beta_i \beta_{K-i-1}$: $|\psi_{2N}\rangle$ is pair product invariant.

Case $\Leftarrow$: also by induction on $n$.

Base case: for $n=2$, the pair product invariance of $|\psi_4\rangle$ is the equality $\alpha_1 \alpha_2 = \alpha_0 \alpha_3$, which implies that $|\psi_4\rangle$ is separable.

Induction step: assume the $\Leftarrow$ property holds for $2, 3, \ldots n$ qubits. Let $|\psi_{2^{n+1}}\rangle = |\psi_{2N}\rangle$ be a state with no zero amplitude and which is pair product invariant:

$$|\psi_{2N}\rangle = \sum_{i=0}^{2N-1} \beta_i |i\rangle$$

Let $\lambda$ be such that $\beta_1 = \lambda \beta_0$ (there always exists such a $\lambda$, since no $\beta_i$ is zero). Since lemma 1 applies to $|\psi_{2N}\rangle$, the equality $\beta_{2(i+1)-1} \beta_{2(i+1)} = \beta_{2(i+1)-2} \beta_{2(i+1)+1}$ is satisfied for $0 \leq i \leq 2^n - 2$. Then, combining this equality with the assumption that $\beta_{2i+1} = \lambda \beta_{2i}$ gives $\beta_{2i+3} \beta_{2i} \beta_{2i+1} = \lambda \beta_{2i+2} \beta_{2i+1} \beta_{2i}$ and proves that $\beta_{2i+3} = \lambda \beta_{2i+2}$ also holds for $0 \leq i \leq 2^n - 2$. Therefore, $\beta_{2i+1} = \lambda \beta_{2i}$ holds for $0 \leq i \leq 2^n - 1$.

Now, it is always possible to take $\delta_0$ and $\delta_1$ with $\delta_1 = \lambda \delta_0$, and such that $|\delta_0|^2 + |\delta_1|^2 = 1$. Then $\frac{\beta_{2i+1}}{\delta_1} = \frac{\beta_{2i}}{\delta_0}$ and $|\psi_{2N}\rangle$ may be written as follows:

$$|\psi_{2N}\rangle = \left( \sum_{i=0}^{N-1} \alpha_i |i\rangle \right) \otimes \begin{pmatrix} \delta_0 \\ \delta_1 \end{pmatrix} = |\psi_N\rangle \otimes \begin{pmatrix} \delta_0 \\ \delta_1 \end{pmatrix} \quad \text{with} \quad \alpha_i = \frac{\beta_{2i+1}}{\delta_1} = \frac{\beta_{2i}}{\delta_0}$$

Since $|\psi_{2N}\rangle$ is pair product invariant, $\beta_{2i}\beta_{2K-2i-1}$ is independent of $i$, for any given $k$, $1 < k \leq n$. As a consequence, $\alpha_i \alpha_{K-i-1} = \dfrac{\beta_{2i}\beta_{2K-2i-1}}{\delta_0 \delta_1}$ is also independent of $i$, which implies that $|\psi_N\rangle$ is pair product invariant. Thus, by induction hypothesis, $|\psi_N\rangle$ is fully separable, and so is also $|\psi_{2N}\rangle$. <span style="float:right">c</span>

## 2.2. GENERAL CASE

States $|\psi_N\rangle = \sum_{i=0}^{N-1} \alpha_i |i\rangle \in \mathcal{H}_N$, where $N = 2^n$, are always considered to be normalized: $\sum_{i=0}^{N-1} |\alpha_i|^2 = 1$.

If $|\psi_2\rangle \in \mathcal{H}_2$ and $|\psi_N\rangle$ is a fully separable state in $\mathcal{H}_N$, then $|\psi_2\rangle \otimes |\psi_N\rangle$ is a fully separable state in $\mathcal{H}_{2N}$. The subset $\mathcal{K}_N$ of all fully separable states in $\mathcal{H}_N$ is defined recursively as follows:

- $\mathcal{K}_2 = \mathcal{H}_2$
- $\mathcal{K}_{2N} = \{ |\psi_2\rangle \otimes |\psi_N\rangle \mid |\psi_2\rangle \in \mathcal{K}_2 \wedge |\psi_N\rangle \in \mathcal{K}_N \}$

With $|\psi_2\rangle = \begin{pmatrix} \delta_0 \\ \delta_1 \end{pmatrix}$, $|\psi_{2N}\rangle \in \mathcal{K}_{2N}$ is $|\psi_{2N}\rangle = \begin{pmatrix} \delta_0 \\ \delta_1 \end{pmatrix} \otimes |\psi_N\rangle$, i.e., for short: $|\psi_{2N}\rangle = \begin{pmatrix} \delta_0 |\psi_N\rangle \\ \delta_1 |\psi_N\rangle \end{pmatrix}$.

An *amplitude abstraction function* $f : \mathcal{H}_N \to \{0,1\}^N$, a set of *well-formed bit strings* $\mathcal{B}_N \subset \{0,1\}^N$, a set of *well-formed states* $\mathcal{V}_N \subset \mathcal{H}_N$ and a family of *zero deletion functions* $g_K : \mathcal{V}_N \to \mathcal{H}_K$, with $K = 2^k$ for $1 \leq k \leq n$, will be useful for characterizing the general case of full separability of $|\psi_N\rangle$, when some amplitudes in $|\psi_N\rangle$ may be zero.

DEFINITION 4. *Amplitude abstraction.* When applied to $|\psi_N\rangle = \sum_{i=0}^{N-1} \alpha_i |i\rangle \in \mathcal{H}_N$, the *amplitude abstraction function* $f : \mathcal{H}_N \to \{0,1\}^N$ yields a bit string $x \in \{0,1\}^N$, with $x = x_0 x_1 \ldots x_{N-1}$ such that, for $0 \leq i \leq N-1$:

- $x_i = 0$ iff $\alpha_i = 0$
- $x_i = 1$ otherwise

DEFINITION 5. *Well-formed bit strings.* The set $\mathcal{B}_N \subset \{0,1\}^N$ of *well-formed bit strings* of length $N$ is defined recursively as follows:

- $\mathcal{B}_2 = \{01, 10, 11\}$
- $\mathcal{B}_{2N} = \{0^N x, x0^N, xx \mid x \in \mathcal{B}_N\}$ where $0^N$ is a string of $N$ 0's.

The number of 1's in $x \in \mathcal{B}_{2N}$ is either the number of 1's in $y \in \mathcal{B}_N$ if $x = 0^N y$ or $x = y0^N$, or twice that number if $x = yy$. Therefore, the number of 1's in $x \in \mathcal{B}_N$ is $K = 2^k$, with $k \leq n$: this number will be denoted by $|x|$. It is also useful to notice that the distributions of the 0's in the string $x$ when $x = yy$ are identical in both halves of $x$.

One may also notice that there are $3^n$ different bit strings in $\mathcal{B}_{2^n}$.

The bit strings in $\mathcal{B}_N$ are said to be "well-formed" because they are amplitude abstractions of fully separable states. This is the purpose of the next lemma:

LEMMA 2. $\forall |\psi_N\rangle \in \mathcal{K}_N : f(|\psi_N\rangle) \in \mathcal{B}_N$

PROOF. By induction on *n*, the number of qubits. □

According to this simple lemma, the bit strings in $\mathcal{B}_N$ tell where the 0's must be for a state $|\psi_N\rangle \in \mathcal{H}_N$ to be fully separable. But not all $|\psi_N\rangle \in \mathcal{H}_N$ such that $f(|\psi_N\rangle) \in \mathcal{B}_N$ are fully separable. There is indeed a set of states $\mathcal{V}_N$, with $\mathcal{K}_N \subset \mathcal{V}_N \subset \mathcal{H}_N$, and such that $\forall |\psi_N\rangle \in \mathcal{V}_N : f(|\psi_N\rangle) \in \mathcal{B}_N$:

DEFINITION 6. *Well-formed states*. The set of *well-formed states* is:

$$\mathcal{V}_N = \{ |\psi_N\rangle \in \mathcal{H}_N \mid f(|\psi_N\rangle) \in \mathcal{B}_N \}$$

When a well-formed state $|\psi_N\rangle \in \mathcal{V}_N$ has a number of zero amplitudes, these are placed correctly for this state to be a candidate to full separability. But this is not enough for it to be fully separable, there must also be conditions satisfied by the non-zero amplitudes of $|\psi_N\rangle$, and by them only. Since $f(|\psi_N\rangle) \in \mathcal{B}_N$, there are $K = 2^k$ non-zero amplitudes in $|\psi_N\rangle$, with $k \leq n$. If $K = 1$, $|\psi_N\rangle$ is not a superposition, hence not entangled: this trivial case will not be considered in what follows. For $K \geq 2$, only the $K$ non-zero amplitudes of $|\psi_N\rangle$ have to be considered. For this, all zero amplitudes will be eliminated from $|\psi_N\rangle$, yielding a state $|\psi_K\rangle \in \mathcal{H}_K$ with no zero amplitudes:

DEFINITION 7. *Zero deletion functions*. For all sets $\mathcal{V}_N$ of well-formed states, there exists a family of *zero deletion functions* $\{ g_K : \mathcal{V}_N \to \mathcal{H}_K \mid K = 2^k, 1 \leq k \leq n \}$ defined as follows:

$$\forall |\psi_N\rangle = \sum_{i=0}^{N-1} \alpha_i |i\rangle \text{ with } |f(|\psi_N\rangle)| = K \text{ and } K \geq 2,$$

$$g_K(|\psi_N\rangle) = \sum_{j=0}^{K-1} \alpha'_j |j\rangle, \text{ where } \alpha'_j = \alpha_i \text{ such that } |\{\alpha_l \mid l < i \wedge \alpha_l \neq 0\}| = j$$

The stage is now set for characterizing full separability of $|\psi_N\rangle$ in the general case:

THEOREM 2. Let $|\psi_N\rangle \in \mathcal{H}_N$ be a state with $|f(|\psi_N\rangle)| = K$ and $2 \leq K \leq N$. Then:

$|\psi_N\rangle$ is fully separable ⇔ $|\psi_N\rangle$ is well-formed and $g_K(|\psi_N\rangle)$ is pair product invariant

PROOF. Case ⇒: by induction on *n*, the number of qubits.

Base case: for $n = 2$, if $|\psi_4\rangle$ is fully separable, then $|\psi_4\rangle$ is well formed since $|\psi_4\rangle \in \mathcal{K}_4 \subset \mathcal{V}_4$ by definition of these sets. If $K = 2$, $g_K(|\psi_4\rangle) \in \mathcal{H}_2$ is trivially pair product invariant. If $K = 4$, the full separability of $|\psi_4\rangle$ implies $\alpha_1 \alpha_2 = \alpha_0 \alpha_3$, which is precisely the pair product invariance of $g_K(|\psi_4\rangle) \in \mathcal{H}_4$.

Induction step: assume the $\Rightarrow$ property holds for 2, 3, ... $n$ qubits. Let $|\psi_{2^{n+1}}\rangle = |\psi_{2N}\rangle$ be a state fully separable into a product of $n+1$ component qubit states, i.e. $|\psi_{2N}\rangle \in \mathcal{K}_{2N}$. Then, by definition of well-formedness, $|\psi_{2N}\rangle$ is indeed well-formed: $f(|\psi_{2N}\rangle) \in \mathcal{B}_{2N}$. Being fully separable, $|\psi_{2N}\rangle$ can in particular be separated into:

$$|\psi_{2N}\rangle = \begin{pmatrix} \delta_0 \\ \delta_1 \end{pmatrix} \otimes |\psi_N\rangle = \begin{pmatrix} \delta_0 |\psi_N\rangle \\ \delta_1 |\psi_N\rangle \end{pmatrix} \text{ where } \begin{pmatrix} \delta_0 \\ \delta_1 \end{pmatrix} \in \mathcal{H}_2 \text{ is normalized}$$

If $\delta_0 = 0$, since $\delta_1 = 1$ and $|f(|\psi_{2N}\rangle)| = |f(|\psi_N\rangle)| = K$ in this case, then $g_K(|\psi_{2N}\rangle) = g_K(|\psi_N\rangle)$: by induction hypothesis on $|\psi_N\rangle$, $g_K(|\psi_{2N}\rangle)$ is pair product invariant. Same reasoning when $\delta_1 = 0$.

If neither $\delta_0$ nor $\delta_1$ is zero, and since $|f(|\psi_{2N}\rangle)| = 2|f(|\psi_N\rangle)| = 2K$ in this case:

$$g_{2K}(|\psi_{2N}\rangle) = \begin{pmatrix} \delta_0 \\ \delta_1 \end{pmatrix} \otimes g_K(|\psi_N\rangle) = \begin{pmatrix} \delta_0 g_K(|\psi_N\rangle) \\ \delta_1 g_K(|\psi_N\rangle) \end{pmatrix}$$

Then, it is easy to verify that the induction hypothesis on $|\psi_N\rangle$, i.e. the pair product invariance of $g_K(|\psi_N\rangle)$, implies the pair product invariance of $g_{2K}(|\psi_{2N}\rangle)$.

Case $\Leftarrow$: also by induction on $n$.

Base case: for $n = 2$, $|\psi_4\rangle$ being well-formed means that $f(|\psi_4\rangle) \in \{0011, 1100, 1010, 0101\}$ when $K = 2$, and that $f(|\psi_4\rangle) = 1111$ when $K = 4$. Then, when $K = 2$, $\alpha_1 \alpha_2 = \alpha_0 \alpha_3 = 0$, and when $K = 4$, the pair product invariance of $g_K(|\psi_4\rangle) \in \mathcal{H}_4$ is $\alpha_1 \alpha_2 = \alpha_0 \alpha_3$: in both cases, $|\psi_4\rangle$ is indeed fully separable.

Induction step: assume the $\Leftarrow$ property holds for 2, 3, ... $n$ qubits. $|\psi_{2N}\rangle$ is well-formed and, for some $K$, $g_K(|\psi_{2N}\rangle)$ is pair product invariant. The $2N$ amplitudes of $|\psi_{2N}\rangle$ can be divided in two halves: the first $N$ amplitudes will be viewed as $\gamma_0 |\varphi_N\rangle$ where $|\varphi_N\rangle \in \mathcal{H}_N$ and the $N$ remaining ones as $\gamma_1 |\chi_N\rangle$ where $|\chi_N\rangle \in \mathcal{H}_N$. The only purpose of the complex coefficients $\gamma_0$ and $\gamma_0$ is to keep $|\psi_{2N}\rangle$ normalized: $|\gamma_0|^2 + |\gamma_1|^2 = 1$, since both $|\varphi_N\rangle$ and $|\chi_N\rangle$ are themselves normalized. This will be summarized visually with the use of the (abusive) notation:

$$|\psi_{2N}\rangle = \begin{pmatrix} \gamma_0 |\varphi_N\rangle \\ \gamma_1 |\chi_N\rangle \end{pmatrix}$$

The well-formedness of $|\psi_{2N}\rangle$ distinguishes three possible distributions of its zero amplitudes:

(i) $f(|\varphi_N\rangle) = 0^N$ and $f(|\chi_N\rangle) \neq 0^N$ with $f(|\chi_N\rangle) \in \mathcal{B}_N$. In this case, $\gamma_1 = 1$, $|f(|\psi_{2N}\rangle)| = |f(|\chi_N\rangle)| = K$ and $g_K(|\psi_{2N}\rangle) = g_K(|\chi_N\rangle)$, which implies that $g_K(|\chi_N\rangle)$ is pair product invariant. Thus, by induction hypothesis, $|\chi_N\rangle$ is fully separable, and so is also $|\psi_{2N}\rangle$ since $|\psi_{2N}\rangle = \begin{pmatrix} 0 \\ 1 \end{pmatrix} \otimes |\chi_N\rangle$.

(ii) $f(|\varphi_N\rangle) \neq 0^N$ and $f(|\chi_N\rangle) = 0^N$ with $f(|\varphi_N\rangle) \in \mathcal{B}_N$. Same reasoning as for case (i).

(iii) $f(|\varphi_N\rangle) = f(|\chi_N\rangle) \neq 0^N$ with $f(|\varphi_N\rangle), f(|\chi_N\rangle) \in \mathcal{B}_N$. In this case, $\gamma_0 \neq 0$, $\gamma_1 \neq 0$ and $|f(|\psi_{2N}\rangle)| = 2|f(|\varphi_N\rangle)| = 2K$. By theorem 1, since $g_{2K}(|\psi_{2N}\rangle)$ has no zero amplitude and is pair product invariant, it is fully separable:

$$g_{2K}(|\psi_{2N}\rangle) = \begin{pmatrix} \delta_0 \\ \delta_1 \end{pmatrix} \otimes |\omega_K\rangle = \begin{pmatrix} \delta_0 |\omega_K\rangle \\ \delta_1 |\omega_K\rangle \end{pmatrix} \text{ for } \begin{pmatrix} \delta_0 \\ \delta_1 \end{pmatrix} \in \mathcal{H}_2 \text{ with } \delta_0, \delta_1 \neq 0$$

and $|\omega_K\rangle \in \mathcal{H}_K$ fully separable and with no zero amplitude.

$f(|\varphi_N\rangle) = f(|\chi_N\rangle)$ means that the positions of the 0's within $|\varphi_N\rangle$ are the same as the positions of the 0's within $|\chi_N\rangle$. It is always possible to find $|\psi_N\rangle \in \mathcal{H}_N$ having all its 0's at the same positions as those of both $|\varphi_N\rangle$ and $|\chi_N\rangle$ and such that $g_K(|\psi_N\rangle) = |\omega_K\rangle$. Therefore $\gamma_0 |\varphi_N\rangle = \delta_0 |\psi_N\rangle$ and $\gamma_1 |\chi_N\rangle = \delta_1 |\psi_N\rangle$, hence:

$$|\psi_{2N}\rangle = \begin{pmatrix} \delta_0 |\psi_N\rangle \\ \delta_1 |\psi_N\rangle \end{pmatrix} = \begin{pmatrix} \delta_0 \\ \delta_1 \end{pmatrix} \otimes |\psi_N\rangle$$

where $|\psi_N\rangle$, a kind of "common divisor" of $|\varphi_N\rangle$ and $|\chi_N\rangle$, is well-formed by construction and is such that $g_K(|\psi_N\rangle) = |\omega_K\rangle$ is pair product invariant by theorem 1, since it is fully separable and with no zero amplitude. In conclusion, the induction hypothesis applies to $|\psi_N\rangle$: $|\psi_N\rangle$ is fully separable and so is also $|\psi_{2N}\rangle$. □

## 3. *p-q* SEPARABILITY

Given integers $p$ and $q$ such that $p + q = n$, $|\psi_N\rangle$ is *p-q* separable iff it can be factorized into a tensor product:

$$|\psi_N\rangle = |\psi_P\rangle \otimes |\psi_Q\rangle$$

where $|\psi_P\rangle \in \mathcal{H}_P$ and $|\psi_Q\rangle \in \mathcal{H}_Q$, with $P = 2^p$ and $Q = 2^q$ (i.e. the two subsystems are composed of $p$ and $q$ qubits respectively). This can be visualized as follows:

$$|\psi_N\rangle = \begin{pmatrix} \alpha_0 \\ \vdots \\ \alpha_{kQ+r} \\ \vdots \\ \alpha_{N-1} \end{pmatrix} = \begin{pmatrix} \beta_0 \\ \vdots \\ \beta_k \\ \vdots \\ \beta_{P-1} \end{pmatrix} \otimes \begin{pmatrix} \gamma_0 \\ \vdots \\ \gamma_r \\ \vdots \\ \gamma_{Q-1} \end{pmatrix} = \begin{pmatrix} \beta_0 \gamma_0 \\ \vdots \\ \beta_0 \gamma_{Q-1} \\ \vdots \\ \beta_k \gamma_r \\ \vdots \\ \beta_{P-1} \gamma_0 \\ \vdots \\ \beta_{P-1} \gamma_{Q-1} \end{pmatrix} \text{ where } |\psi_P\rangle = \begin{pmatrix} \beta_0 \\ \vdots \\ \beta_{P-1} \end{pmatrix} \text{ and } |\psi_Q\rangle = \begin{pmatrix} \gamma_0 \\ \vdots \\ \gamma_{Q-1} \end{pmatrix}.$$

In other words, $|\psi_N\rangle = \begin{pmatrix} \alpha_0 \\ \vdots \\ \alpha_{N-1} \end{pmatrix}$ is *p-q* separable iff $\forall k \in [0, P-1]$ and $\forall r \in [0, Q-1]$, $\alpha_{kQ+r} = \beta_k \gamma_r$.

This defines a structure for the $N$ amplitudes in $|\psi_N\rangle$. There are $P$ groups of $Q$ amplitudes each: $\alpha_{kQ+r} = \beta_k \gamma_r$ is amplitude $r$ ($r \in [0, Q-1]$) in group $k$ ($\forall k \in [0, P-1]$). This means that, if $\beta_k = 0$, all amplitudes in group $k$ are equal to zero, and that the distributions of zeros within all groups obtained with $\beta_k \neq 0$ are the same and are identical to the distribution of zeros within $|\psi_Q\rangle$.

THEOREM 3. Let $|\psi_N\rangle \in \mathcal{H}_N$, with $N = PQ$, be a state such that for some $i_0 = k_0 Q + r_0 \in [0, N-1]$, $\alpha_{i_0} \neq 0$ and $\forall i < i_0$, $\alpha_i = 0$. Then:

$$|\psi_N\rangle \text{ is } p\text{-}q \text{ separable} \iff \forall k \in [k_0+1, P-1], \forall r \in [r_0+1, Q-1], \alpha_{k_0 Q + r_0} \alpha_{kQ+r} = \alpha_{k_0 Q + r} \alpha_{kQ + r_0}$$

PROOF. Case $\Rightarrow$: $|\psi_N\rangle$ $p$-$q$ separable means that $\forall k \in [0, P-1]$ and $\forall r \in [0, Q-1]$, $\alpha_{kQ+r} = \beta_k \gamma_r$. Then: $\forall k \in [k_0+1, P-1], \forall r \in [r_0+1, Q-1]: \alpha_{k_0 Q + r_0} \alpha_{kQ+r} = \beta_{k_0} \gamma_{r_0} \beta_k \gamma_r = \beta_{k_0} \gamma_r \beta_k \gamma_{r_0} = \alpha_{k_0 Q + r} \alpha_{kQ+r_0}$.

Case $\Leftarrow$: given $|\psi_N\rangle$ where $\alpha_{k_0 Q + r_0}$ is the first non zero amplitude, and such that $\forall k \in [k_0+1, P-1], \forall r \in [r_0+1, Q-1], \alpha_{k_0 Q + r_0} \alpha_{kQ+r} = \alpha_{k_0 Q + r} \alpha_{kQ+r_0}$, the problem is to prove that there exist quantum states $|\psi_P\rangle$ and $|\psi_Q\rangle$ such that $|\psi_N\rangle = |\psi_P\rangle \otimes |\psi_Q\rangle$. The proof goes by finding $|\psi_P\rangle$ and $|\psi_Q\rangle$ such that $\forall k \in [0, P-1], \forall r \in [0, Q-1], \alpha_{kQ+r} = \beta_k \gamma_r$, where

$$|\psi_N\rangle = \begin{pmatrix} \alpha_0 \\ : \\ \alpha_{N-1} \end{pmatrix} \text{ is given, and where } |\psi_P\rangle = \begin{pmatrix} \beta_0 \\ : \\ \beta_{P-1} \end{pmatrix}, |\psi_Q\rangle = \begin{pmatrix} \gamma_0 \\ : \\ \gamma_{Q-1} \end{pmatrix} \text{ are normalized quantum states.}$$

Instances of such states $|\psi_P\rangle$ and $|\psi_Q\rangle$ can be obtained by choosing, $\forall k \in [0, P-1], \forall r \in [0, Q-1]$:

$$\gamma_{r_0} \text{ such that } \gamma_{r_0} \gamma_{r_0}^* = 1 \Bigg/ \left( 1 + \frac{\sum_{i=i_0+1}^{(k_0+1)Q-1} \alpha_i \alpha_i^*}{\alpha_{i_0} \alpha_{i_0}^*} \right), \quad \gamma_r = \frac{\gamma_{r_0} \alpha_{k_0 Q + r}}{\alpha_{i_0}}, \text{ and } \beta_k = \frac{\alpha_{kQ+r_0}}{\gamma_{r_0}}.$$

This choice implies that $|\psi_N\rangle = |\psi_P\rangle \otimes |\psi_Q\rangle$. Indeed, $\forall k \in [k_0+1, P-1]$:

$$\forall r \in [r_0+1, Q-1], \beta_k \gamma_r = \frac{\alpha_{kQ+r_0}}{\gamma_{r_0}} \frac{\gamma_{r_0} \alpha_{k_0 Q + r}}{\alpha_{i_0}} = \frac{\alpha_{k_0 Q + r} \alpha_{kQ+r_0}}{\alpha_{i0}}$$

$$= \frac{\alpha_{k_0 Q + r_0} \alpha_{kQ+r}}{\alpha_{i0}} \text{ by the hypothesis of case } \Leftarrow$$

$$= \alpha_{kQ+r}$$

For $r = r_0$, $\beta_k \gamma_{r_0} = \alpha_{kQ+r_0}$ by definition of $\beta_k$ and $\gamma_{r_0}$.

$\forall r \in [0, r_0-1]$, $\alpha_{k_0 Q + r} = 0$. Then, $\gamma_k = 0$ by its definition, and $\alpha_{kQ+r} = 0$ by the hypothesis of case $\Leftarrow$. Hence: $\beta_k \gamma_r = \alpha_{kQ+r}$.

For the other values of $k$: $k = k_0 \Rightarrow \beta_{k_0} \gamma_r = \alpha_{k_0 Q + r}$ by definition of $\beta_{k_0}$ and $\gamma_r$, and, $\forall k \in [0, k_0 - 1]$, $\alpha_{kQ+r} = \beta_k \gamma_r$ since $\alpha_{kQ+r} = 0$ and $\beta_k = 0$ because $\alpha_{kQ+r_0} = 0$.

Finally, since $|\psi_N\rangle = |\psi_P\rangle \otimes |\psi_Q\rangle$, $|\psi_P\rangle$ and $|\psi_Q\rangle$ are normalized quantum states:

$$\||\psi_Q\rangle\|^2 = \sum_{r=0}^{Q-1} \gamma_r \gamma_r^* = \sum_{r=0}^{r_0-1} \gamma_r \gamma_r^* + \gamma_{r_0} \gamma_{r_0}^* + \sum_{r=r_0+1}^{Q-1} \gamma_r \gamma_r^*$$

$$= \gamma_{r_0} \gamma_{r_0}^* + \sum_{r=r_0+1}^{Q-1} \gamma_r \gamma_r^* \text{ because } \beta_{k_0} \neq 0 \wedge \forall r < r_0, \alpha_{k_0 Q + r} = 0 \Rightarrow \forall r < r_0, \gamma_r = 0$$

$$= \gamma_{r_0} \gamma_{r_0}^* (1 + \sum_{r=r_0+1}^{Q-1} \frac{\alpha_{k_0 Q + r} \alpha_{k_0 Q + r}^*}{\alpha_{i_0} \alpha_{i_0}^*}) = 1$$

$$\||\psi_P\rangle\|^2 = \frac{\||\psi_N\rangle\|^2}{\||\psi_Q\rangle\|^2} = 1$$

Hence $|\psi_N\rangle$ is *p-q* separable. ▫

## 4. A SHORT NOTE ON COMPLEXITY

Given a state $|\psi_N\rangle$ and an integer $p$, the test for *p-q* separability based upon theorem 3 requires $(P-1)(Q-1) = N - P - Q + 1$ comparisons of $2(N - P - Q + 1)$ products of two amplitudes in the worst case (i.e. when $\alpha_0 \neq 0$).

The analysis of the situation is a little more complicated for full separability in the general case (theorem 2), i.e. when some amplitudes may be equal to zero, because, in addition to checking pair product invariance among non-zero amplitudes, it is necessary to check that $|\psi_N\rangle$ is well-formed with respect to the distribution of its zero amplitudes.

If there are $K = 2^k$ non-zero amplitudes in $|\psi_N\rangle$, checking pair product invariance among them amounts to verifying that, for $l \in [1, k]$, the products $\alpha_i \alpha_{L-i-1}$ are constant for $i \in [0, L-1]$ where, $\forall l, L = 2^l$. Since the case $l = 1$ is trivially satisfied and since, by symmetry, all products $\alpha_i \alpha_{L-i-1}$ in the range $i \in \left[\frac{L}{2}, L-1\right]$ are already taken care of in the range $i \in \left[0, \frac{L}{2} - 1\right]$, this results in $\sum_{l=2}^{k}(2^{l-1} - 1) = K - k - 1$ comparisons among $K - 2$ products of two amplitudes. The worst case is $K = N$: the number of products and the number of comparisons for checking pair product invariance are both of the order of *N*.

Let $z$ be the list of all integers $i$ sorted in increasing order and such that $\alpha_i = 0$, where $\alpha_i$ is an amplitude in $|\psi_N\rangle$. Let $low[z,m]$ be a function which returns the sorted sublist of all integers $i$ where $i$ is an element of $z$ such that $i < m$, and $high[z,m]$ be a function which returns the sublist of all integers $i \bmod m$, where $i$ is an element of $z$ such that $m \leq i$. According to the definition of well-formed states, $|\psi_N\rangle$ is well-formed if $wf?\left[low\left[z,\frac{N}{2}\right], high\left[z,\frac{N}{2}\right], N\right]$ returns *true*, given the function $wf?$ defined as follows, where $|l|$ denotes the length of a list $l$:

$$wf?[l,h,N] = \text{If } N=2 \;\; : \;\; true$$

$$\text{If } |l| > |h| \;\; : \;\; \{\text{If } |l| = \frac{N}{2} \;\; : \;\; wf?\left[low\left[h,\frac{N}{4}\right], high\left[h,\frac{N}{4}\right], \frac{N}{2}\right]. \text{ Otherwise : } false\}$$

$$\text{If } |l| < |h| \;\; : \;\; \{\text{If } |h| = \frac{N}{2} \;\; : \;\; wf?\left[low\left[l,\frac{N}{4}\right], high\left[l,\frac{N}{4}\right], \frac{N}{2}\right]. \text{ Otherwise : } false\}$$

$$\text{Otherwise : } \;\; \{\text{If } l = h \;\; : \;\; wf?\left[low\left[l,\frac{N}{4}\right], high\left[l,\frac{N}{4}\right], \frac{N}{2}\right]. \text{ Otherwise : } false\}$$

The recursion depth of $wf?\left[low\left[z,\frac{N}{2}\right], high\left[z,\frac{N}{2}\right], N\right]$ is $log_2(|z|)$, where $|z|$ is the number of zeros among the amplitudes in $|\psi_N\rangle$. At each level, this algorithm performs either 2 tests for comparing $|l|$ and $|h|$, followed by at most $\frac{|z|}{2}$ comparisons of integers for checking whether $l$ and $h$ are equal, or 1 test followed by scanning a list of length less than $\frac{|z|}{2}$. Its complexity is, in the worst case:

$$2\, log_2(|z|) + \frac{1}{2} \sum_{i=0}^{\lfloor log_2|z| \rfloor} \frac{|z|}{2^i}, \text{ which means } |z| + \Theta(log_2(|z|)) \text{ comparisons.}$$

Noting that, in any case, $|z| \leq N-1$, checking that $|\psi_N\rangle$ is well-formed will cost of the order of $N$ comparisons. Finally, the total complexity of the test for full separability is of the order of $N$ products and $2N$ comparisons.

## 5. CONCLUSION

The literature shows that a lot has been achieved about understanding entangled pure and mixed states of bi-partite quantum systems but, beyond three qubit systems, entanglement is, for the most part, *terra incognita*. This paper is a contribution to studying separability criteria for *n*-partite systems, while staying on the safe ground of pure states only, like other recent attempts to analyzing entanglement beyond two- or three-partite quantum systems [17][18].

Further questions are suggested by the approach presented here:

(i) the search for values of *p* and *q*, while minimizing the number of adequate permutations of qubits for deciding whether or not $|\psi_N\rangle$ is *p*-*q* separable; this requires a fine analysis of which subsets remain invariant within sets of pair product equalities when two qubits are swapped.

(ii) the relation of pair-product invariance, as defined for full separability, but also of its generalization to *p*-*q* separability, with other invariants and with measures of entanglement (e.g. concurrence, see [17]).

(iii) the definition of classes of quantum states for which the characterization of *p*-*q* or *p*-*q*-*r*-... separability could rely on less complex criteria; a first example of such a class could be a set of states defined by stabilizers and closed under a specified set of quantum operations.